\documentclass[journal]{IEEEtran}
\usepackage{graphicx}
\usepackage{subcaption}
\usepackage[cmex10]{amsmath, mathtools}
\usepackage{array,multirow,graphicx,subfig,fleqn,float}
\usepackage{booktabs}
\usepackage{tabularx}
\usepackage{xcolor}
\usepackage{amssymb}
\usepackage{comment}
\usepackage{makecell}

\newcolumntype{Y}{>{\centering\arraybackslash}X}

\begin{document}

\title{Ultra-Lightweight Network for Ship-Radiated Sound Classification on Embedded Deployment}

\author{Sangwon Park, Dongjun Kim, Sung-Hoon Byun, Sangwook Park,~\IEEEmembership{Member,~IEEE,}
\thanks{Manuscript received December 1, 2025}
\thanks{Sangwon Park, Dongjun Kim, and Sangwook Park are with the Department of Electronic and Semiconductor Engineering, Gangneung-Wonju National University, Gangwon-do, 25457, Republic of Korea, e-mail: spark2@gwnu.ac.kr.(corresponding author), Sung-Hoon Byun is with Department of Ocean and Maritime Digital Technology Research, Korea Research Institute of Ships \& Ocean Engineering, Daejeon, 34103, Republic of Korea}
}

\markboth{Journal of \LaTeX\ Class Files,~Vol.~xx, No.~x, September~2023}%
{Shell \MakeLowercase{\textit{et al.}}: Bare Demo of IEEEtran.cls for Journals}

\maketitle

\begin{abstract}
This letter presents ShuffleFAC, a lightweight acoustic model for ship-radiated sound classification in resource-constrained maritime monitoring systems. ShuffleFAC integrates Frequency-Aware convolution into an efficiency-oriented backbone using separable convolution, point-wise group convolution, and channel shuffle, enabling frequency-sensitive feature extraction with low computational cost. Experiments on the DeepShip dataset show that ShuffleFAC achieves competitive performance with substantially reduced complexity. In particular, ShuffleFAC ($\gamma=16$) attains a macro F1-score of $71.45\pm1.18~\%$ using $39~K$ parameters and $3.06~M$ MACs, and achieves an inference latency of $6.05\pm0.95$ ms on a Raspberry Pi. Compared with MicroNet0, it improves macro F1-score by 1.82 \% while reducing model size by 9.7× and latency by 2.5×. These results indicate that ShuffleFAC is suitable for real-time embedded UATR.
\end{abstract}


\IEEEpeerreviewmaketitle

\section{Introduction}
\IEEEPARstart{W}{ith} the rapid growth of maritime traffic, Underwater Acoustic Target Recognition (UATR), particularly ship-radiated sound classification, has become increasingly important for maritime surveillance, traffic monitoring, and marine environmental protection~\cite{gamage2023comprehensive, bjorno2017underwater}. In practical deployments, sensing platforms such as buoy networks and/or fixed hydrophone nodes operate continuously and autonomously under harsh ocean conditions. Reliable UATR on such platforms therefore requires acoustic models that are robust to ambient noise, reverberation, and diverse operating conditions. Recent deep learning–based approaches trained on large-scale real-world recordings consistently outperform traditional methods based on hand-crafted acoustic features~\cite{irfan2021deepship,4590381,9929447}.

\begin{figure}[!t]
\centering
\includegraphics[width=0.7\columnwidth]{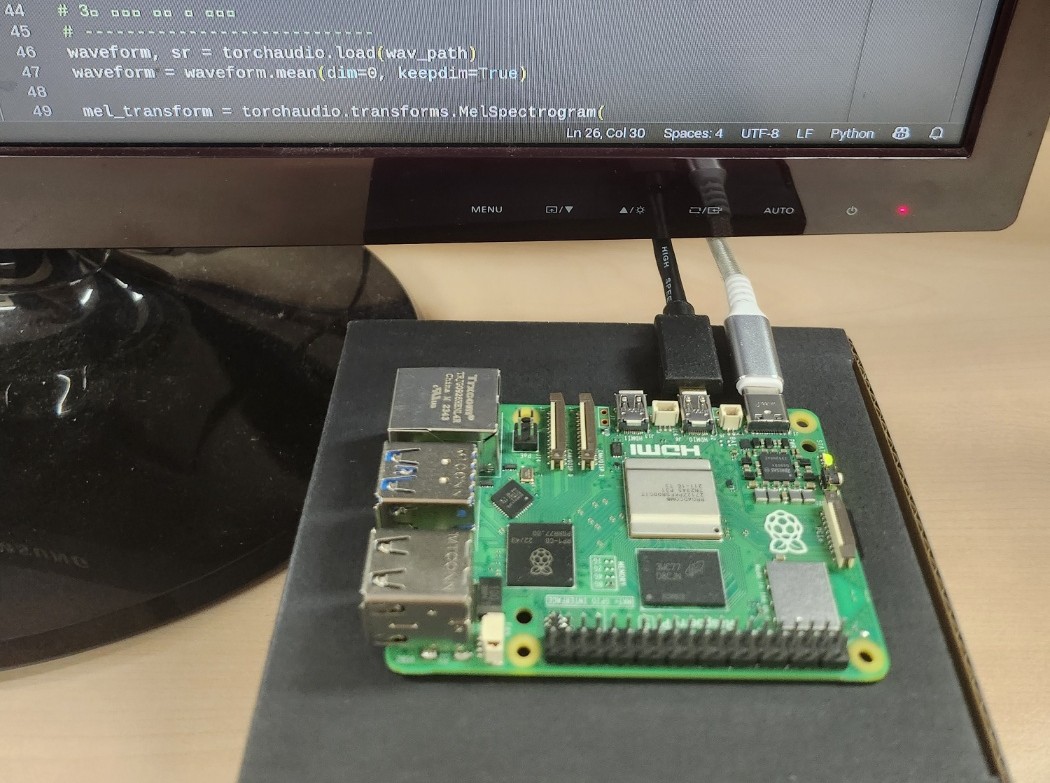}
\caption{Raspberry Pi 5: Resource-constraint platform}
\label{fig:raspberry}
%\vspace{-0.5cm}
\end{figure}

However, on-device UATR remains challenging due to strict resource constraints. Power is typically supplied by batteries and energy harvesting, limiting the computational budget and duty cycle of onboard processing. Moreover, embedded hardware often relies on low-power CPUs or microcontrollers without high-performance GPUs, which restricts feasible model complexity. Memory and storage constraints further require careful control of model size. Despite these limitations, high recognition accuracy is essential because false alarms and missed detections can directly impact maritime safety and security. While Convolutional Neural Network (CNN)- and Transformer-based acoustic models achieve strong performance in high-resource settings, their parameter counts and compute demand often make direct deployment impractical on embedded UATR platforms~\cite{irfan2021deepship, feng2022transformer}.

%% Section 1-3&4. cutting down version
%To mitigate this mismatch, lightweight architectures such as MobileNet~\cite{howard2017mobilenets,sandler2018mobilenetv2}, ShuffleNet~\cite{zhang2018shufflenet,ma2018shufflenet}, and MicroNet~\cite{li2021micronet} reduce complexity via separable/group/factorized convolution. Nevertheless, these models often rely on shift-invariant convolution, which can be suboptimal for spectrogram-like inputs where frequency position matters, and they may incur non-negligible deployment overhead due to tensor manipulations not reflected in MAC counts. This letter proposes ShuffleFAC, which combines frequency-aware feature extraction with an efficiency-oriented backbone designed to reduce both arithmetic cost and deployment overhead. Experiments on DeepShip demonstrate that ShuffleFAC provides a favorable trade-off among classification performance, model size, MACs, and embedded inference latency.
To address these constraints, resource-efficient network architectures are introduced for edge deployment on mobile and embedded platforms (Fig.~\ref{fig:raspberry}). MobileNet employs separable convolution, decomposing a standard convolution into depthwise and point-wise operations~\cite{howard2017mobilenets,sandler2018mobilenetv2}. Building on this idea, ShuffleNet adopts point-wise group convolution and channel shuffle to reduce computation while enabling cross-group information exchange~\cite{zhang2018shufflenet,ma2018shufflenet}. More recently, MicroNet further reduces complexity by introducing micro-factorized separable convolution~\cite{li2021micronet}. Nevertheless, these models largely rely on shift-invariant convolution, which can be suboptimal for spectrogram-like inputs where frequency-axis position is informative~\cite{song2025frequency}. Moreover, their practical efficiency on embedded devices can be affected by tensor-manipulation overhead (e.g., memory allocation and reshape/split operations) that is excluded in Multiply-Accumulate (MAC) counts~\cite{lee2026hiddencostsinferencedeep}.

This letter proposes \emph{ShuffleFAC}, which combines frequency-aware feature extraction with an efficiency-oriented backbone to reduce both arithmetic cost and deployment overhead. Experiments on the \emph{DeepShip} dataset, which contains underwater ship-radiated sound recordings, demonstrate that the proposed model achieves a favorable trade-off among classification performance, model size, MACs, and measured embedded inference latency. Specifically, with the channel scaling factor set to 
$\gamma=16$, ShuffleFAC attains a macro F1-score of $71.45\%$ using $39~K$ parameters and $3.06~M$ MACs, and runs in $6.05$ ms on a Raspberry Pi 5. Compared with MicroNet0, it improves macro F1 by $1.82\%$ while reducing model size by $9.7\times$ and inference latency by $2.5\times$.

\section{Related Work}
\subsection{Separable convolution for Lightweight CNNs}
Let $\mathbf{X}\in\mathbb{R}^{C_{in}\times H\times W}$ be an input feature map. A standard $k_h \times k_w$ convolution with $C_{out}$ output channels requires $\mathcal{O}(HWk_hk_wC_{in}C_{out})$ operations and $k_h k_w C_{in}C_{out}$ parameters. Separable convolution reduces this cost by factorizing the operation into a depthwise convolution and a point-wise ($1 \times 1$) convolution, resulting in $\mathcal{O}\big(HW(k_h k_w C_{in}+C_{in}C_{out})\big)$ operations and $C_{in}(k_h k_w + C_{out})$ parameters. This factorization forms the basis of MobileNet-style architectures for efficient on-device inference~\cite{howard2017mobilenets,sandler2018mobilenetv2}.

%% Section II-A-2. before cut-down
%Using this strategy, MobileNet dramatically reduces computation and model size, enabling deployment on mobile devices~\cite{howard2017mobilenets}. ShuffleNet further improves efficiency by adopting point-wise group convolution, which reduces the computational cost to $\mathcal{O}\big(HW(k_h k_w C_{in}+C_{in}C_{out}/g)\big)$ with $g$-groups, and by introducing channel shuffle to enable information exchange across groups~\cite{zhang2018shufflenet}. More recently, MicroNet achieves additional reductions by introducing micro-factorized depthwise and point-wise convolutions. For depthwise convolution, a 2-D spatial kernel is factorized as $p \otimes q^T$, where $p \in \mathbb{R}^{k_h \times 1}$ and $q \in \mathbb{R}^{k_w \times 1}$. For point-wise convolution, the weight matrix is decomposed via an intermediate subspace as $\mathbf{W}=\mathbf{P}\boldsymbol{\phi}\mathbf{Q}$, where $\mathbf{P} \in \mathbb{R}^{C_{out} \times M}$, $\boldsymbol{\phi} \in \mathbb{R}^{M \times M}$, and $\mathbf{Q} \in \mathbb{R}^{M \times C_{in}}$~\cite{li2021micronet}. This micro-factorized separable convolution reduces the computational complexity to $\mathcal{O}\big((H k_h + W k_w +HWC_{out})C_{in}\big)$, making inference more efficient on resource-constrained platforms.
%% Section II-A-2. less version
ShuffleNet improves efficiency by using point-wise group convolution, reducing point-wise cost by approximately $1/g$ with $g$ groups, and applies channel-shuffle to enable cross-group information exchange~\cite{zhang2018shufflenet, ma2018shufflenet}. MicroNet further reduces complexity through micro-factorization of spatial kernels and channel projections~\cite{li2021micronet}. Specifically, the 2-D spatial kernel used in depthwise convolution is factorized as $p \otimes q^T$, where $\otimes$ denotes outer product, $p \in \mathbb{R}^{k_h \times 1}$, and $q \in \mathbb{R}^{k_w \times 1}$. The weight matrix of the point-wise convolution is decomposed via an intermediate subspace as $\mathbf{W}=\mathbf{P}\boldsymbol{\phi}\mathbf{Q}^T$, where $\mathbf{P} \in \mathbb{R}^{C_{in} \times C_{int}}$, $\boldsymbol{\phi} \in \mathbb{R}^{C_{int} \times C_{int}}$, and $\mathbf{Q} \in \mathbb{R}^{C_{out} \times C_{int}}$~\cite{li2021micronet}. With these factorizations, micro-factorized separable convolution reduces the MAC count to $\mathcal{O}\big((H k_h + W k_w)C_{in}+HWC_{int}(C_{in}+C_{out})\big)$ and the number of parameters to $C_{in}(k_h + k_w) + C_{int}(C_{in}+C_{out})$. While these approaches reduce arithmetic cost, practical embedded latency can still be influenced by tensor manipulations (e.g. tensor split, slice, copy) and memory movement beyond MAC counts.

%% initial version
%Previous studies have proposed various efficient neural network architectures optimized for on-device environments in computer vision tasks.
%\textbf{MobileNet~\cite{howard2017mobilenets}} reduces the number of parameters and computational cost by employing depthwise separable convolutions, making it well-suited for devices with limited memory such as smartphones or embedded systems. \textbf{MobileNetV2~\cite{sandler2018mobilenetv2}} employs not only depthwise separable convolution but also inverted residual with linear bottle neck.
%\textbf{ShuffleNet~\cite{zhang2018shufflenet}}, a follow-up to MobileNet, further improves efficiency by incorporating pointwise group convolution, depthwise convolution, and a channel shuffle operation. This design maintains model accuracy while significantly reducing computational overhead. \textbf{ShuffleNetV2~\cite{ma2018shufflenet}} further improves practical efficiency by introducing design guidelines that reduce memory access cost and operator fragmentation, thereby achieving faster real-world inference.
%More recently, \textbf{MicroNet~\cite{li2021micronet}} achieves substantial reductions in FLOPs by introducing micro-factorized depthwise and point convolutions, enabling highly compact and efficient models for resource-constrained platforms.

\vspace{-0.25cm}
\subsection{Frequency Adaptive Convolution}
%% Section II-B-1. before cut-down
%In spectrogram based inputs, the same local pattern can convey different semantics depending on its frequency band. Accordingly, frequency adaptive convolution incorporates explicit frequency-position information so that the effective filtering can vary along the frequency axis.
%% Section II-B-1. after revision
In time-frequency representations, the same local pattern can convey different semantics depending on its frequency band. Frequency-adaptive convolution incorporates frequency-position information so that the effective filtering can vary along the frequency axis.

%% Section II-B-2. before cut-down
%Frequency DYnamic convolution (FDY) achieves frequency adaptivity by employing multiple frequency-conditioned kernels. Typically, an auxiliary branch produces frequency-dependent kernels from the input, and the output is obtained by combining $k$ candidate kernels~\cite{nam2022frequency}. This design increases the computational load because convolution must be performed with multiple kernels. The complexity scales as $\mathcal{O}\big(HWk_hk_wC_{in}C_{out}k\big)$, and the parameter count increased to $k_h k_w C_{out}C_{in}k$.
%On the other hand, FA convolution provides a lower-overhead alternative by injecting frequency information through positional encoding. FA convolution augments the feature map with a frequency encoding vector and applies adaptive, channel-independent scaling to match the encoding magnitude to the input feature statistics prior to convolution~\cite{song2025frequency}. Because the frequency conditioning is realized through lightweight modulation rather than multiple kernels, FA convolution introduces only marginal extra MACs and parameters relative to the underlying backbone.
%(i.e. standard and/or separable convolution).
Frequency dynamic convolution typically achieve frequency adaptivity by combining multiple frequency-conditioned kernels~\cite{nam2022frequency}. Because it evaluates and aggregates responses from multiple kernels, its computational and parameter costs scale with the number of basis kernels. In contrast, Frequency Aware Convolution (FAC) injects frequency-position information using a positional encoding and lightweight channel-wise modulation prior to convolution~\cite{song2025frequency}. FAC achieve frequency sensitivity without requiring multiple kernel evaluations, making it attractive for embedded deployment.

%% initial version
%Recently, frequency adaptive convolution based acoustic model leads the state-of-the-art in the field of Sound Event Detection in air.
%In this study, an acoustic model is required for ship-radiated noise classification. In recent years, acoustic deep neural networks have demonstrated strong performance, particularly in sound event detection. Frequency Dynamic Convolution (FDY)~\cite{nam2022frequency} applies frequency-adaptive kernels to impose frequency dependency on 2D convolution. However, FDY is computationally heavy and unsuitable for embedded systems. To reduce computational cost, Frequency Aware Convolution (FAC)~\cite{song2025frequency} was proposed. FAC extracts features sensitive to the frequency positions of time–frequency patterns by adding a frequency encoding vector to the CNN input. This encoding vector is adaptively scaled in a channel-independent manner to match the input amplitude while preserving frequency-dependent characteristics.

\section{Proposed Method}
When designing an acoustic model for embedded UATR, the top priority is to achieve both computational efficiency and strong classification accuracy. To this end, the proposed model is designed by combining frequency-aware feature extraction with an efficiency-oriented backbone.

\vspace{-0.25cm}
\subsection{Frequency Adaptive Separable Convolution Module}
\begin{figure}[!t]
\centering
\includegraphics[width=0.9\columnwidth]{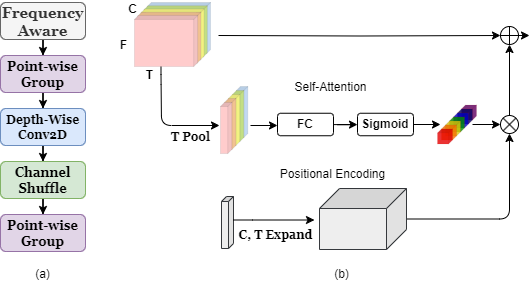}
\caption{Illustrations of (a) Frequency Adaptive Separable Convolution module, (b) Frequency aware pipeline}
\label{fig:block_diagram}
\vspace{-0.25cm}
\end{figure}

%Fig.~\ref{fig:block_diagram}(a) illustrates a diagram of Frequency-Adaptive Separable Convolution (FASC) module, which combines a Frequency Aware (FA) block with a separable-convolution with channel shuffle. As shown in Fig.~\ref{fig:block_diagram}(b), the FA block injects frequency-position information by adding a learnable frequency encoding that is modulated by channel-wise gating. This design enables frequency-sensitive feature extraction with only mild overhead, consisting mainly of element-wise operations and a small matrix multiplication in the FC layer of the self-attention branch.
Fig.~\ref{fig:block_diagram}(a) illustrates a diagram of Frequency-Adaptive Separable Convolution (FASC) module, which integrates a frequency-aware (FA) block into an efficiency-oriented separable-convolution. The FA block injects frequency-position information through a learnable positional encoding (Fig.~\ref{fig:block_diagram}(b)). Specifically, the encoding is broadcast along the time and channel dimensions to match the input feature-map shape and is modulated by channel-wise gates produced by a lightweight self-attention branch (temporal pooling, a Fully-Connected (FC) layer, and sigmoid activation). The resulting channel-weighted positional bias is then added to the input feature map. This design enables frequency-sensitive feature extraction with only mild overhead, since it relies primarily on element-wise operations and a matrix multiplication rather than additional convolutional kernels. Because the injected bias is constant over time, the FA block is particularly effective for quasi-stationary signals, which is consistent with ship-radiated sound characteristics.

A ShuffleNet-style separable block requires approximately $\mathcal{O}\big(HWk_hk_w C_{in}^{2} + 2HWC_{in}^{2}/g\big)$ operations when the output channel is doubled ($C_{out}=2C_{in}$). In this setting, the depthwise term typically dominates the total complexity because $k_hk_w > 2/g$. To mitigate this issue, channel compression is applied prior to the depthwise convolution. Specifically, after the FA block, a point-wise group convolution (with $g_{1}=2$) reduces the channel dimension by a factor of two, followed by a depthwise convolution for spatial feature extraction at the reduced channels. Channel shuffle is then applied to enable cross-group information exchange, and a second point-wise group convolution (with $g_{2}=2$) expands the channels to twice the input channel for the subsequent stage. Notably, increasing the number of groups reduces the MACs of point-wise group convolution, but it can also introduce additional tensor-manipulation overhead that degrades practical latency on embedded CPUs.

%\subsection{Proposed Model Architecture}
%We study ship-radiated sound classification on low-power embedded systems. Therefore, not only a lightweight architecture but also an effective acoustic model is required. To address the limited frequency sensitivity of the vanilla convolution, frequency adaptive convolution is required.
%Accordingly, we employ FAC~\cite{song2025frequency} instead of FDY~\cite{nam2022frequency} to reduce computational cost. The computational costs, measured in Multiply-Accumulate operations, for FDY, and FAC are compared below. FDY utilizes $N$ basis kernels to generate frequency-adaptive weights. 
%However, its implementation requires performing convolution operations individually for each of the $N$ basis kernels before aggregating the results~\cite{nam2022frequency}. Consequently, the computational cost of FDY scales linearly with $N$, as shown in Eq.~(\ref{eq:MACs_comparison}). In contrast, FAC \cite{song2025frequency} achieves frequency adaptation without such redundant convolutions, maintaining a cost comparable to vanilla convolution.

\begin{comment}
\begin{align}
\label{eq:MACs_comparison}
\text{MACs}_\text{FDY} &\approx N \times (K^{2} \times H_{\text{out}} \times W_{\text{out}} \times C_{\text{in}} \times C_{\text{out}}) \\
\text{MACs}_\text{FAC} &\approx 1 \times (K^{2} \times H_{\text{out}} \times W_{\text{out}} \times C_{\text{in}} \times C_{\text{out}})
\end{align}
\end{comment}

%where $C_{\text{in}}$ and $C_{\text{out}}$ denote the number of input and output channels, respectively, and $K$ represents the kernel size. $H_{\text{out}}$ and $W_{\text{out}}$ correspond to the output feature map dimensions. $N$ denotes the number of basis kernels in FDY \cite{song2025frequency} (typically $N=4$), which acts as a direct multiplier to the computational load.
%As illustrated in Fig.~\ref{fig:block_diagram} a, Frequency Aware explicitly injects frequency positional information into the CNN input. In Fig.~\ref{fig:block_diagram} a, positional encoding vectors are learnable parameters. During model training, the proposed model learns frequency-dependent weights.

\begin{table}[!t]
    \centering
    \caption{Proposed Model Architecture}
    \label{tab:architecture}
    \renewcommand{\arraystretch}{1.3} 
    \small 
    \renewcommand{\tabularxcolumn}[1]{m{#1}}
    
    \begin{tabularx}{\columnwidth}{
        >{\centering\arraybackslash}m{1.2cm} 
        Y 
        >{\centering\arraybackslash}m{1.8cm} 
    }
        \toprule
        {Stage} & {Configuration} & {Output Shape} \\
        \midrule
        Input & Log-mel spectrogram & $1 \times 128 \times 24$ \\
        \midrule
        
        \makecell{Channel \\ Expansion} & \makecell{FA, Conv2D ($\gamma$),  ReLU, \\ BN, AvgPool ($2\times2$)} & $\gamma \times 64 \times 12$ \\
        \midrule
        
        \multirow{12}{*}{\centering \makecell{Feature \\ Embedding}} 
          & \makecell{FASC ($2\gamma$), ReLU, BN, \\ AvgPool ($2\times2$)} & $2\gamma \times 32 \times 6$ \\ 
          \cmidrule{2-3}
          & \makecell{FASC ($4\gamma$), ReLU, BN, \\ AvgPool ($1\times2$)} & $4\gamma \times 16 \times 6$ \\ 
          \cmidrule{2-3}
          & \makecell{FASC ($8\gamma$), ReLU, BN, \\ AvgPool ($1\times2$)} & $8\gamma \times 8 \times 6$ \\ 
          \cmidrule{2-3}
          & \makecell{FASC ($8\gamma$), ReLU, BN, \\ AvgPool ($1\times2$)} & $8\gamma \times 4 \times 6$ \\ 
          \cmidrule{2-3}
          & \makecell{FASC ($8\gamma$), ReLU, BN, \\ AvgPool ($1\times2$)} & $8\gamma \times 2 \times 6$ \\ 
          \cmidrule{2-3}
          & \makecell{FASC ($8\gamma$), ReLU, BN, \\ AvgPool ($1\times2$)} & $8\gamma \times 1 \times 6$ \\
          \midrule
          Classifier & Global AvgPool, Linear (8$\gamma$, 4) & $4$ \\
        \bottomrule
    \end{tabularx}
\end{table}

\vspace{-0.25cm}
\subsection{ShuffleFAC Architecture}
For embedded UATR, a lightweight acoustic model is designed by stacking the proposed FASC modules with ReLU activation, Batch Normalization (BN), and Average pooling (AvgPool), as summarized in Table~\ref{tab:architecture}. Given a log-Mel spectrogram with 128-frequency bands and 24-frames as the input tensor $\mathbf{X} \in \mathbb{R}^{1 \times128 \times 24}$, the first stage applies the FA block and a standard convolution from~\cite{song2025frequency} to perform initial channel expansion to $\gamma$. Subsequent layers employ the FASC module for efficient frequency-adaptive feature extraction, where the channel is progressively doubled up to the fourth stage, while later stages keep the channel dimension fixed.

Average pooling is used to down-sample the frequency dimension throughout the network, while the time dimension is reduced mainly in the first two stages. This channel and pooling configuration reduces the overall computational cost of separable convolution. Moreover, unlike MobileNetV2~\cite{sandler2018mobilenetv2} and ShuffleNetV2~\cite{ma2018shufflenet}, the proposed architecture avoids residual and parallel paths, thereby reducing tensor-manipulation overhead that can degrade practical latency on embedded processors. Finally, global average pooling is applied to the final feature map, and a linear layer projects the pooled feature vector to the target class space.

\vspace{-0.25cm}
\subsection{Model Training}
The model parameters are optimized using mini-batch stochastic gradient descent by minimizing the Cross-Entropy (CE) loss 
\begin{equation}
\label{ce}
\mathcal{L}_{\text{CE}}(y, \hat{y}) = - \frac{1}{N} \sum_{n=1}^{N} \sum_{c=1}^{C} y_{n,c} \log(\hat{y}_{n,c}),
\end{equation}
where $N$ and $C$ denote the number of samples in a mini-batch and the number of target classes, respectively. Here, $y_{n,c}$ is the one-hot ground-truth label and $\hat{y}_{n,c}$ is the model's prediction that the $n$th sample belongs to class $c$. During the training, Adam optimizer with a learning rate of $0.001$ is applied under the setting of batch size 48 ($N=48$) and a maximum of 200 epochs\footnote{Online available: https://github.com/KNU-LMAP/ShuffleFAC}.

%In this work, we set the number of groups to 2 to balance computational efficiency and feature representation. In Table~\ref{tab:architecture}, $\gamma$ denotes the number of channels. For example, in shuffleFAC 32–256, the value 32 corresponds to $\gamma_{1}$, while 256 corresponds to $\gamma_{4}$. The intermediate channel sizes, $\gamma_{2}$ and $\gamma_{3}$, are set to 64 and 128, respectively. The model is optimized using the Adam optimizer with a learning rate of $1\times10^{-3}$. The batch size is set to 48, and the proposed model is trained for 200 epochs. The Cross-Entropy loss used for  model training is defined as Eq.~(\ref{ce}).

%Where $N$ is the total number of data samples in the batch, and $C$ is the total number of classes to be classified. $y_{i,c}$ is an element of the One-Hot Encoded vector, taking a value of $1$ if the $i$-th sample's true label belongs to class $c$, and $0$ otherwise. $\hat{y}_{i,c}$ denotes the model's predicted probability that the $i$-th sample belongs to class $c$.

\begin{table*}[ht!]
\centering
\caption{Comparison of performance each models}
\label{tab:comp_ratio}
\resizebox{\textwidth}{!}{
    % REMOVED scalebox{0.05} because it makes the table microscopic. 
    % resizebox{\textwidth} handles the scaling correctly.
    \begin{tabular}{@{}lcccccc@{}}
    \toprule
    \textbf{Model} & \textbf{Accuracy (\%)} & \textbf{F1-score (\%)} & \textbf{\# of Params} & \textbf{MACs} & \textbf{$t_{\mathrm{inf}}$ (ms)} & \textbf{$E_{\mathrm{est}}$ ($\mu$Wh)} \\
    \midrule
    $^\ast$FAC & $72.85\pm1.90$ & $73.12\pm2.14$ & $2.45~\text{M}$ & $129.25~\text{M}$ & $13.97\pm1.13$ & $34.92\pm2.83$ \\
    \midrule
    $^\ast$SCAE & $72.37\pm1.18$ & $72.59\pm1.25$ & $1.53~\text{M}$ & $560.03~\text{M}$ & $45.22\pm0.66$ & $113.05\pm1.66$ \\
    \midrule
    MobileNet & $70.55\pm0.64$ & $70.85\pm0.74$ & $3.22~\text{M}$ & $40.99~\text{M}$ & $14.69\pm3.05$ & $36.71\pm7.63$ \\
    MobileNetV2 & $68.92\pm0.02$ & $69.42\pm0.20$ & $2.25~\text{M}$ & $28.70~\text{M}$ & $13.95\pm0.78$ & $34.88\pm$1.96 \\
    \midrule
    ShuffleNet & $69.17\pm3.37$ & $69.46\pm3.48$ & $919~\text{K}$ & $10.48~\text{M}$ & $13.70\pm0.56$ & $34.25\pm1.41$\\
    ShuffleNetV2 & $69.58\pm1.25$ & $69.91\pm0.13$ & $1.26~\text{M}$ & $10.73~\text{M}$ & $11.02\pm0.49$ & $27.55\pm1.23$ \\
    \midrule
    MicroNet 3 & $70.40\pm0.77$ & $70.81\pm0.78$ & $1.60~\text{M}$ & $3.32~\text{M}$ & $23.05\pm0.92$ & $57.64\pm2.31$ \\
    MicroNet 2 & $69.55\pm1.13$ & $70.11\pm1.09$ & $1.37~\text{M}$ & $2.27~\text{M}$ & $23.67\pm1.50$ & $59.18\pm3.77$ \\
    MicroNet 1 & $68.75\pm1.09$ & $69.33\pm1.07$ & $818~\text{K}$ & $1.21~\text{M}$ & $16.23\pm1.77$ & $40.58\pm4.44$ \\
    MicroNet 0 & $69.04\pm0.46$ & $69.63\pm0.50$ & $379~\text{K}$ & $0.65~\text{M}$ & $15.13\pm1.26$ & $37.82\pm3.16$\\
    \midrule
    ShuffleFAC ($\gamma=64$) & $70.26\pm0.52$ & $70.17\pm0.79$ & $546~\text{K}$ & $34.64~\text{M}$ & $10.86\pm0.84$ & $27.16\pm2.11$ \\
    ShuffleFAC ($\gamma=32$) & $71.44\pm1.71$ & $71.66\pm1.74$ & $143~\text{K}$ & $9.85~\text{M}$ & $7.49\pm1.31$ & $18.73\pm3.29$ \\
    ShuffleFAC ($\gamma=16$) & $71.31\pm1.34$ & $71.45\pm1.18$ & $39~\text{K}$ & $3.06~\text{M}$ & $6.05\pm0.95$ & $15.14\pm2.37$ \\
    ShuffleFAC ($\gamma=8$) & $69.15\pm1.46$ & $69.38\pm1.67$ & $11~\text{K}$ & $1.06~\text{M}$ & $5.48\pm0.70$ & $13.71\pm1.77$ \\
    \bottomrule
    \multicolumn{7}{r}{\scriptsize{$^\ast$our implementation}} \\
    \end{tabular}
}
\end{table*}

\section{Experiments}
\subsection{Database and Pre-processing}
\label{subsec:data}
Experiments are conducted on the DeepShip dataset, which contains approximately 47 h of real-world underwater recordings from 265 ships across four vessel categories~\cite{irfan2021deepship}. To construct non-overlapping subsets, the original recordings are split at the file level into training, validation, and test sets with a ratio of 7:1:2. Each recording is resampled to 16 kHz and segmented into 3-s audio clips without overlaid. This procedure yields approximately 56,000 clips. Because segmentation is performed after the file-level split, all clips from a given recording are assigned to a single subset, preventing train–test contamination and better reflecting performance on unseen recordings. Each 3-s clip is converted into a log-Mel spectrogram by computing the short-time Fourier transform with a 256-ms window and 128-ms hop, taking the magnitude, applying a 128-channel Mel filterbank, and then performing log-amplitude scaling.

%% initial version
%In the experiments, we used Deepship dataset~\cite{irfan2021deepship}, which consists of 47 h and 4 min of real world underwater recordings of 265 different ships belong to four classes. We divided each audio file in the DeepShip dataset into training, validation, and test sets with a ratio of 7:1:2. All audio clips are 3 seconds in length and were resampled to 16kHz. A 4096 Short-Time Fourier Transform is applied to convert the audio signals into spectrograms, which are then transformed into Mel-spectrograms using Mel filter banks. 

\subsection{Evaluation Metrics and Experimental Settings}
This study compares the proposed model with representative lightweight architectures adapted to the DeepShip-based UATR task by making minor modifications to their input/output layers and pooling settings. Specifically, FAC, Separable Convolution based Autoencoder (SCAE), MobileNet, ShuffleNet, and MicroNet are considered as baseline models. All models are trained and evaluated on a high-resource workstation equipped with an NVIDIA RTX A5000 GPU.

In the assessment, two types of metrics are used to evaluate classification accuracy and efficiency. For classification, micro \emph{Accuracy} and macro \emph{F1-score} are reported. Accuracy is computed at the clip level, while macro F1-score is obtained by averaging the classwise F1-scores. The F1-score is the harmonic mean of precision and recall, where a correct prediction contributes one true positive and an incorrect prediction contributes one false positive and one false negative. A small gap between Accuracy and F1-score indicates more balanced performance across targets. Each assessment is repeated at least three times, and the results are reported as the mean and standard deviation.

To assess efficiency, the number of trainable parameters and MACs are reported as measures of model size and computational complexity. In addition, inference latency of each model is measured on a Raspberry Pi 5 with 8 GB RAM to capture practical overhead from tensor manipulation that is excluded in MAC counts (Fig.~\ref{fig:raspberry}). Finally, the energy consumption per inference is estimated from the measured inference time under an assumed CPU utilization of 90\% as $E_{\mathrm{est}}=0.9P_{\mathrm{cpu}} \times t_{\mathrm{inf}}/3600$, where $P_{\mathrm{cpu}}$ is the CPU peak power in Watts and $t_{\mathrm{inf}}$ is the inference time in seconds.

\subsection{Results}

Table~\ref{tab:comp_ratio} summarizes the classification performance and model complexity of all models. For most architectures, micro Accuracy and macro F1-score are close, indicating relatively balanced performance across the vessel categories rather than strong bias toward a subset of classes. As two benchmarks, FAC and SCAE achieve strong classification performance as Accuracy: $72.85\pm1.90$\% and $72.37\pm1.18$\%, respectively. However both are less suitable for embedded deployment due to their high computational cost and latency. In particular, SCAE exhibits the largest MACs and the highest inference time on Raspberry Pi, which is consistent with its Xception-style design that stacks multiple separable convolutions with residual connections, which increases both operations and runtime overhead~\cite{chollet2017xception}.

Conventional lightweight architectures reduce model complexity at the expense of recognition performance. MobileNet decreases the MACs substantially relative to FAC ($129.25~M$ $\rightarrow$ $40.99~M$) with moderate performance degradation, while MobileNetV2 further reduces MACs and parameters but shows additional degradation in both Accuracy and F1. ShuffleNet and ShuffleNetV2 operate in the $\sim10~M$ MACs and achieve improved embedded latency as $13.70\pm0.56$ ms and $11.02\pm0.49$ ms, respectively. But their performance remains below MobileNet. MicroNet variants achieve extremely low MACs (down to $0.65~M$ for MicroNet0), but their classification performance still remains at the level of the MobileNets. Also, their measured latency still remains relatively high about $15\sim23$ ms, suggesting that practical runtime on embedded platform is influenced by factors beyond MACs, such as memory access and tensor-operator overhead. This issue is analyzed in more detail in the next section.

The proposed ShuffleFAC models provide a favorable accuracy–efficiency trade-off. Among the proposed variants, ShuffleFAC ($\gamma=16$) achieves the best position in the accuracy–complexity space (Accuracy: $71.31\pm1.34$\%, F1: $71.45\pm1.18$\%) with only $39~K$ parameters and $3.06~M$ MACs, while achieving an inference latency of $6.05\pm0.95$ ms and an estimated energy of $15.14\pm2.37~\mu$Wh on the Raspberry Pi. Notably, ShuffleFAC attains competitive accuracy with substantially lower model size and computational cost than conventional lightweight architectures. Across configurations, $\gamma=8$ yields the lowest latency and energy, whereas increasing the width to $\gamma=32$ achieves comparable accuracy; further scaling to $\gamma=64$ increases compute and latency without improving performance. Overall, these results indicate that ShuffleFAC is well suited for real-time UATR on resource-constrained platforms.

%We compared the classification performance of several neural networks on the DeepShip dataset. As presented in Table~\ref{tab:comp_ratio}, all models demonstrate reasonable performance on Deepship dataset. Among them the ShuffleNet achieves the highest overall performance, with an Accuracy of $73.22\pm0.45\%$ and a Macro F1-score of $73.56\pm0.46\%$ with 919K parameters.

%However, 919K parameters remain relatively large for deployment on low-power embedded platforms. In contrast, the proposed ShuffleFAC~16–128 achieves competitive performance an Accuracy of $71.31\pm1.71\%$ and a Macro F1-score of $71.45\pm1.18\%$ with 39K parameters. This represents a substantial reduction in model size while maintaining strong performance. Such efficiency stems from the architectural design, where pointwise group convolution significantly reduces computational cost, and the channel shuffle operation compensates for the limited cross-group information flow by restoring inter-channel interactions.

%As illustrated in Fig.~\ref{fig:acc_vs_param}, ShuffleFAC shows a degradation in accuracy when the model size exceeds 200k parameters. In contrast, under the 200k-parameter constraint, ShuffleFAC achieves superior performance compared to ShuffleNet, highlighting its effectiveness in ultra-lightweight settings.

% --- Figure 4 placed here (before Figure 3) because it is referenced in text first ---
\begin{figure}[t!]
\centering
\includegraphics[width=0.7\columnwidth]{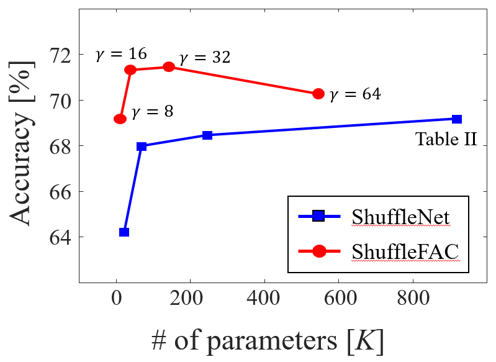}
\caption{Classification accuracy versus model size}
\label{fig:acc_vs_param}
\vspace{-0.25cm}
\end{figure}

\section{Discussions}
\subsection{Comparison with Scaled ShuffleNet}
ShuffleNet variants with different channel dimensions are compared to the corresponding ShuffleFAC variants under the same training and test protocol. Fig.~\ref{fig:acc_vs_param} plots micro Accuracy versus the number of trainable parameters, where the upper-left region indicates a more favorable accuracy–model-size trade-off. As the ShuffleNet backbone is scaled down, its Accuracy degrades noticeably, indicating that the baseline becomes increasingly capacity-limited at small model sizes. In contrast, ShuffleFAC consistently achieves higher Accuracy than ShuffleNet at comparable parameter budgets. This improvement suggests that injecting frequency-position information helps compensate for the reduced capacity of lightweight backbones by enabling more discriminative feature extraction from time–frequency representations.

\vspace{-0.25cm}
\subsection{Embedded Inference Overhead Beyond MACs}
Unlike a high-resource workstation equipped with parallel processors, an embedded platform typically relies on a single CPU to execute the entire inference pipeline. As a result, the processor must handle not only core arithmetic (e.g., convolutions) but also auxiliary operations such as parameter loading, memory allocation, and tensor manipulation (e.g., copy, slice, and concatenation). MAC counts, however, account only for the core multiply–accumulate operations.

%MicroNet0 achieves the lowest MACs by factorizing spatial kernels via outer products and compressing point-wise weights through an intermediate channel subspace, which is implemented using FC based operations. Profiling on a Raspberry Pi shows that MicroNet0 spends approximately 2.27 ms on tensor operations, accounting for about 15\% of the total inference time, yet this overhead is excluded from MAC counts. When such tensor-manipulation overhead is substantial, measured embedded latency is not necessarily proportional to MACs, consistent with prior deployment-oriented observations (e.g.,~\cite{ma2018shufflenet}) and related work on practical inference efficiency [ref].

MicroNet0 achieves the lowest MACs by adopting micro-factorizing separable convolution. Profiling on a Raspberry Pi shows that MicroNet0 spends approximately 2.27 ms on tensor operations, accounting for about 15\% of the total inference time, yet this overhead is excluded from MAC counts. When such tensor-manipulation overhead is non-negligible, measured embedded latency is not necessarily proportional to MACs, which is consistent with prior deployment-oriented observations (e.g.,~\cite{ma2018shufflenet}) and related work on practical inference efficiency~\cite{lee2026hiddencostsinferencedeep}. In contrast, ShuffleFAC ($\gamma=16$) is designed to minimize tensor manipulation and spends only about 0.53 ms (approximately 9\% of total time) on tensor operations. Because most of its runtime (about 82\%) is dominated by core arithmetic, MACs provide a more reliable proxy for comparing computational complexity across ShuffleFAC variants.

\vspace{-0.25cm}
\subsection{Comparison with Recent Studies}
Recent studies report UATR classification accuracies of around 96\%~\cite{feng2022transformer,deng2025advancing}. In those studies, segmented clips are randomly shuffled and then split into training and test sets. Under such segment-level partitioning, clips from the same source recording can appear in multiple subsets, so the model may be evaluated on test clips that are highly similar to recordings observed during training. In contrast, this study adopts a stricter recording-level split as described in Section~\ref{subsec:data}. Consequently, all test clips originate from recordings that are never observed during training, providing a more reliable assessment of generalization to unseen ship-radiated sounds. In addition, this study prioritizes lightweight model design for resource-constrained deployment, which can further limit peak classification performance compared with heavier architectures.

%In \cite{irfan2021deepship}, each segment was processed using a frame length of 250 ms with a hop size of 64 ms. In our experiment, however, we utilized a frame length of 256 ms with a 128 ms hop size. Consequently, the SCAE model \cite{irfan2021deepship} exhibited performance results differing from those originally reported.
%Recent studies \cite{feng2022transformer, deng2025advancing} proposed methods for underwater acoustic target recognition, both achieving classification accuracies exceeding 96\%. These results can be largely attributed to their data partitioning strategies. In \cite{feng2022transformer}, all segments are randomly shuffled prior to being divided into a 70\% training set and a 30\% test set. Similarly, in \cite{deng2025advancing}, segments were randomly shuffled and then allocated to training (70\%), testing (15\%), and validation (15\%) sets. As a result, the data splits in these methods were not totally exclusive at the source recording level, potentially introducing data leakage.

\section{Conclusions}

This letter introduces ShuffleFAC, a lightweight acoustic model for efficient ship-radiated sound classification in resource-constrained maritime monitoring systems. By integrating Frequency-Aware convolution with a channel-shuffling mechanism, the proposed model captures frequency-dependent acoustic patterns while maintaining a compact architecture with reduced computational cost. Experiments on the DeepShip dataset show that ShuffleFAC achieves competitive classification performance with substantially fewer parameters than existing lightweight models. Compared with MicroNet0, ShuffleFAC ($\gamma=16$) improves the macro F1-score by $1.82~\%$ ($69.63 \rightarrow 71.45~\%$), reduces model size by $9.7\times$ ($379\rightarrow39~K$), and achieves $2.5\times$ faster inference ($15.13\rightarrow 6.05~ms$) on Raspberry Pi platform. These results highlight its suitability for real-time deployment. Future work will focus on improving classification performance and validating the proposed model on additional datasets to further assess its generalization capability.

\section*{Acknowledgment}
This research was supported by Basic Science Research Program through the National Research Foundation of Korea (NRF) funded by the Ministry of Science and ICT (RS-2024-00358953) and Korea Research Institute for Defense Technology planning and advancement(KRIT) funded by the Korea government (KRITCT-23-026)

\ifCLASSOPTIONcaptionsoff
  \newpage
\fi

\bibliographystyle{IEEEtran}
\bibliography{bibtex/bib/IEEEabrv}

@article{gamage2023comprehensive,
  title={A comprehensive survey on the applications of machine learning techniques on maritime surveillance to detect abnormal maritime vessel behaviors},
  author={Gamage, Chamali and Dinalankara, Randima and Samarabandu, Jagath and Subasinghe, Akila},
  journal={WMU Journal of Maritime Affairs},
  volume={22},
  number={4},
  pages={447--477},
  year={2023},
  publisher={Springer}
}

@incollection{bjorno2017underwater,
  title={Underwater acoustic measurements and their applications},
  author={Bj{\o}rn{\o}, L},
  booktitle={Applied underwater acoustics},
  pages={889--947},
  year={2017},
  publisher={Elsevier}
}

@article{irfan2021deepship,
  title={DeepShip: An underwater acoustic benchmark dataset and a separable convolution based autoencoder for classification},
  author={Irfan, Muhammad and Jiangbin, ZHENG and Ali, Shahid and Iqbal, Muhammad and Masood, Zafar and Hamid, Umar},
  journal={Expert Systems with Applications},
  volume={183},
  pages={115270},
  year={2021},
  publisher={Elsevier}
}

@inproceedings{zhang2018shufflenet,
  title={Shufflenet: An extremely efficient convolutional neural network for mobile devices},
  author={Zhang, Xiangyu and Zhou, Xinyu and Lin, Mengxiao and Sun, Jian},
  booktitle={Proceedings of the IEEE conference on computer vision and pattern recognition},
  pages={6848--6856},
  year={2018}
}

@inproceedings{song2025frequency,
  title={Frequency-aware convolution for sound event detection},
  author={Song, Tao and Zhang, Wenwen},
  booktitle={International Conference on Multimedia Modeling},
  pages={415--426},
  year={2025},
  organization={Springer}
}

@inproceedings{li2021micronet,
  title={Micronet: Improving image recognition with extremely low flops},
  author={Li, Yunsheng and Chen, Yinpeng and Dai, Xiyang and Chen, Dongdong and Liu, Mengchen and Yuan, Lu and Liu, Zicheng and Zhang, Lei and Vasconcelos, Nuno},
  booktitle={Proceedings of the IEEE/CVF International conference on computer vision},
  pages={468--477},
  year={2021}
}

@inproceedings{ma2018shufflenet,
  title={Shufflenet v2: Practical guidelines for efficient cnn architecture design},
  author={Ma, Ningning and Zhang, Xiangyu and Zheng, Hai-Tao and Sun, Jian},
  booktitle={Proceedings of the European conference on computer vision (ECCV)},
  pages={116--131},
  year={2018}
}

@inproceedings{sandler2018mobilenetv2,
  title={Mobilenetv2: Inverted residuals and linear bottlenecks},
  author={Sandler, Mark and Howard, Andrew and Zhu, Menglong and Zhmoginov, Andrey and Chen, Liang-Chieh},
  booktitle={Proceedings of the IEEE conference on computer vision and pattern recognition},
  pages={4510--4520},
  year={2018}
}

@article{howard2017mobilenets,
  title={Mobilenets: Efficient convolutional neural networks for mobile vision applications},
  author={Howard, Andrew G and Zhu, Menglong and Chen, Bo and Kalenichenko, Dmitry and Wang, Weijun and Weyand, Tobias and Andreetto, Marco and Adam, Hartwig},
  journal={arXiv preprint arXiv:1704.04861},
  year={2017}
}

@article{feng2022transformer,
  title={A transformer-based deep learning network for underwater acoustic target recognition},
  author={Feng, Sheng and Zhu, Xiaoqian},
  journal={IEEE Geoscience and Remote Sensing Letters},
  volume={19},
  pages={1--5},
  year={2022},
  publisher={IEEE}
}

@article{deng2025advancing,
  title={Advancing underwater acoustic target recognition in low-SNR environments with UATR-DIFF-transformer},
  author={Deng, Shuwen and Hong, Feng},
  journal={Ocean Engineering},
  volume={341},
  pages={122668},
  year={2025},
  publisher={Elsevier}
}

@article{nam2022frequency,
  title={Frequency dynamic convolution: Frequency-adaptive pattern recognition for sound event detection},
  author={Nam, Hyeonuk and Kim, Seong-Hu and Ko, Byeong-Yun and Park, Yong-Hwa},
  journal={arXiv preprint arXiv:2203.15296},
  year={2022}
}

@INPROCEEDINGS{9929447,

  author={Lian, Zixu and Wu, Tianshu},

  booktitle={2022 IEEE 6th Advanced Information Technology, Electronic and Automation Control Conference (IAEAC )}, 

  title={Feature Extraction of Underwater Acoustic Target Signals Using Gammatone Filterbank and Subband Instantaneous Frequency}, 

  year={2022},

  volume={},

  number={},

  pages={944-949},

  doi={10.1109/IAEAC54830.2022.9929447}}

@INPROCEEDINGS{4590381,
  author={Li Xin-xin and Yang Shi-e and Yu Ming},
  booktitle={2008 International Conference on Neural Networks and Signal Processing}, 
  title={Feature extraction from underwater signals using wavelet packet transform}, 
  year={2008},
  volume={},
  number={},
  pages={400-405},
  doi={10.1109/ICNNSP.2008.4590381}}

@inproceedings{chollet2017xception,
  title={Xception: Deep learning with depthwise separable convolutions},
  author={Chollet, Fran{\c{c}}ois},
  booktitle={Proceedings of the IEEE conference on computer vision and pattern recognition},
  pages={1251--1258},
  year={2017}
}

@inproceedings{lee2026hiddencostsinferencedeep,
      title={Hidden costs for inference with deep network on embedded system devices}, 
      author={Chankyu Lee and Woohyun Choi and Sangwook Park},
      booktitle={Proceedings of the IEEE conference on consumer electronics},
      year={2025},
      eprint={2601.01698},
      archivePrefix={arXiv},
      primaryClass={cs.CC},
      url={https://arxiv.org/abs/2601.01698}, 
}

\end{document}